\documentclass[onecolumn,published,a4paper]{quantumarticle}
\usepackage{amsmath}
\usepackage{amssymb}
\usepackage{amsthm}
\usepackage{amsfonts}
\usepackage[caption=false]{subfig}
\usepackage[colorlinks]{hyperref}
\usepackage[all]{hypcap}
\usepackage{tikz}
\usepackage{relsize}
\usepackage{color,soul}
\usepackage[utf8]{inputenc}
\usepackage{capt-of}
\usepackage{mathtools}
\usepackage[numbers,sort&compress]{natbib}
\usepackage{float}
\usepackage[section]{placeins}
\usepackage{listings}
\usepackage[T1]{fontenc}  
\usetikzlibrary{decorations.pathreplacing}
\usepackage{braket}

\DeclareFixedFont{\ttb}{T1}{txtt}{bx}{n}{5}
\DeclareFixedFont{\ttm}{T1}{txtt}{m}{n}{5}
\definecolor{deepblue}{rgb}{0,0,0.5}
\definecolor{deepred}{rgb}{0.6,0,0}
\definecolor{deepgreen}{rgb}{0,0.5,0}
\newcommand\cppstyle{\lstset{
language=C++,
basicstyle=\ttm,
otherkeywords={uint8_t, __m256i, size_t, ASSERT_TRUE, EXPECT_TRUE, TEST, BENCHMARK},
keywordstyle=\ttb\color{deepblue},
emphstyle=\ttb\color{deepblue},
stringstyle=\color{deepgreen},
commentstyle=\fontfamily{txtt}\selectfont\color{gray},
showstringspaces=false,
literate={*}{{\char42}}1
         {-}{{\char45}}1
}}
\lstnewenvironment{cpp}[1][]
{\cppstyle\lstset{#1}}{}

\newcommand\pythonstyle{\lstset{
language=python,
basicstyle=\ttm,
morekeywords={assert,as,echo},
keywordstyle=\ttb\color{deepblue},
emphstyle=\ttb\color{deepblue},
stringstyle=\color{deepgreen},
commentstyle=\fontfamily{txtt}\selectfont\color{gray},
showstringspaces=false,
literate={*}{{\char42}}1
         {-}{{\char45}}1
}}
\lstnewenvironment{python}[1][]
{\pythonstyle\lstset{#1}}{}

\lstdefinestyle{stimcircuit}{
    language=python,
    basicstyle=\linespread{0.8}\tiny\ttfamily,
    upquote=true,
    stepnumber=1,
    numbersep=8pt,
    showstringspaces=false,
    breaklines=true,
    frame=single,
    aboveskip=1.5em,
    belowskip=1.5em,
    commentstyle=\fontfamily{txtt}\color{gray},
    classoffset=1,
    morekeywords={DETECTOR,OBSERVABLE_INCLUDE},
    keywordstyle=\color{deepgreen},
    classoffset=2,
    morekeywords={H,R,MPP,M},
    keywordstyle=\ttb\color{deepblue},
    classoffset=3,
    morekeywords={X_ERROR,DEPOLARIZE2,DEPOLARIZE1},
    keywordstyle=\color{red},
    classoffset=4,
    morekeywords={TICK,SHIFT_COORDS,QUBIT_COORDS},
    keywordstyle=\color{gray}
}

\renewcommand\thesection{\arabic{section}}

\theoremstyle{definition}

\theoremstyle{definition}

\theoremstyle{definition}

\newcommand{\eq}[1]{\hyperref[eq:#1]{Equation~\ref*{eq:#1}}}
\renewcommand{\sec}[1]{\hyperref[sec:#1]{Section~\ref*{sec:#1}}}
\DeclareRobustCommand{\app}[1]{\hyperref[app:#1]{Appendix~\ref*{app:#1}}}
\newcommand{\fig}[1]{\hyperref[fig:#1]{Figure~\ref*{fig:#1}}}
\newcommand{\tbl}[1]{\hyperref[tbl:#1]{Table~\ref*{tbl:#1}}}
\newcommand{\theoremref}[1]{\hyperref[theorem:#1]{Theorem~\ref*{theorem:#1}}}
\newcommand{\definitionref}[1]{\hyperref[definition:#1]{Definition~\ref*{definition:#1}}}

\begin{document}
\title{A Fault-Tolerant Honeycomb Memory}

\date{\today}
\author{Craig Gidney}
\email{craig.gidney@gmail.com}
\affiliation{Google Quantum AI, Santa Barbara, California 93117, USA}

\author{Michael Newman}
\email{mgnewman@google.com}
\affiliation{Google Quantum AI, Santa Barbara, California 93117, USA}

\author{Austin Fowler}
\affiliation{Google Quantum AI, Santa Barbara, California 93117, USA}

\author{Michael Broughton}
\affiliation{Google Quantum AI, Santa Barbara, California 93117, USA}

\begin{abstract}

Recently, Hastings \& Haah introduced a quantum memory defined on the honeycomb lattice \cite{hastings2021dynamically}.  Remarkably, this honeycomb code assembles weight-six parity checks using only two-local measurements.  The sparse connectivity and two-local measurements are desirable features for certain hardware, while the weight-six parity checks enable robust performance in the circuit model.

In this work, we quantify the robustness of logical qubits preserved by the honeycomb code using a correlated minimum-weight perfect-matching decoder. Using Monte Carlo sampling, we estimate the honeycomb code's threshold in different error models, and project how efficiently it can reach the ``teraquop regime'' where trillions of quantum logical operations can be executed reliably.  We perform the same estimates for the rotated surface code, and find a threshold of $0.2\%-0.3\%$ for the honeycomb code compared to a threshold of $0.5\%-0.7\%$ for the surface code in a controlled-not circuit model.  In a circuit model with native two-body measurements, the honeycomb code achieves a threshold of $1.5\% < p <2.0\%$, where $p$ is the collective error rate of the two-body measurement gate - including both measurement and correlated data depolarization error processes.  With such gates at a physical error rate of $10^{-3}$, we project that the honeycomb code can reach the teraquop regime with only $600$ physical qubits.  
\end{abstract}

\maketitle

\section{Introduction}
\label{sec:introduction}

Geometrically local codes have been tremendously successful in designing fault-tolerant quantum memories.  They embed in nearest-neighbor layouts and can be highly robust.  Most famous among these is the surface code \cite{kitaev2003faultbyanyons, fowler2012surfacecodereview}, which only requires $4$-body measurements on a square qubit lattice and boasts one of the highest circuit-level thresholds among all quantum codes - typically estimated to be between $0.5\%-1.0\%$ depending on the specifics of the noise model \cite{stephens2014fault}.

However, we can make things even more local.  When encoding information into a subsystem of the codespace \cite{kribs2005unified, bacon2006operator}, parity constraints can be collected using only $3$- \cite{bravyi2012subsystem, kubica2021single} and even $2$-body \cite{bacon2006operator, suchara2011constructions, bombin2010topological, bombin2012universal} measurements.  This is achieved by decomposing the parity constraints into non-commuting measurements of the unprotected degrees of freedom.  This added locality can simplify quantum error-correction circuits by either sparsifying the physical layout or compactifying the syndrome circuit.  While both of these qualities may be desirable at the hardware-level \cite{chamberland2020topological, chamberland2020triangular, chao2020optimization}, the locality often comes at significant cost to the quality of error-correction.  Intuitively, by releasing degrees of freedom to increase locality, we also collect less information about the errors occurring.  This manifests as high-weight parity constraints that can require a large assembly of local measurements to determine.  In the most extreme case, this can eliminate a threshold altogether \cite{bacon2006operator}.

From this perspective, Kitaev's honeycomb model \cite{kitaev2006anyons} is one of the most alluring candidates for building a quantum code.  This model supports $6$-body operators that both commute with, and can be built from, the non-commuting $2$-local terms of its Hamiltonian.  These operators are primed to serve as low-weight parity constraints for a quantum code requiring only $2$-body measurements.  Unfortunately, as several authors have previously noted, a subsystem code defined directly from these constraints encodes no protected logical qubits \cite{suchara2011constructions, lee2017topological}.

\begin{figure}
    \centering
    \resizebox{0.49\linewidth}{!}{
            \includegraphics{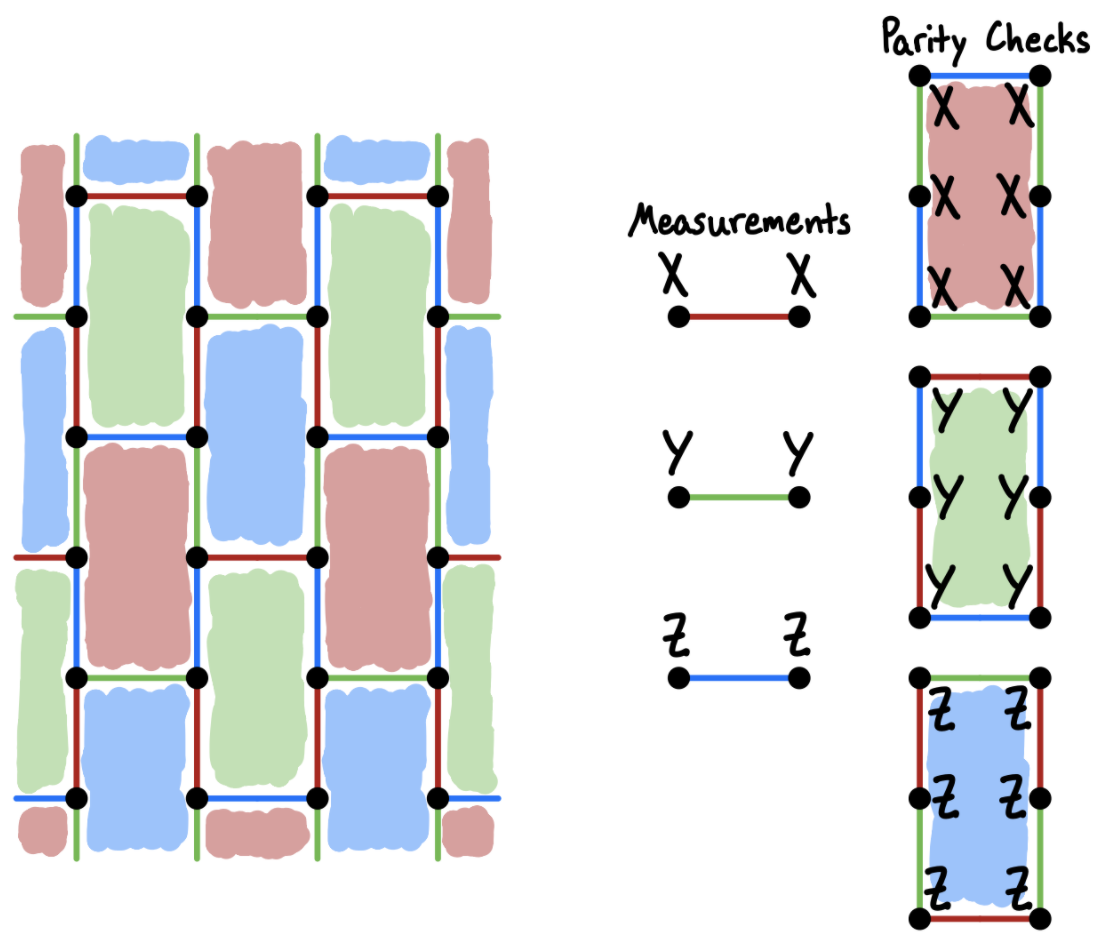}
    }
    \resizebox{\linewidth}{!}{
        \includegraphics{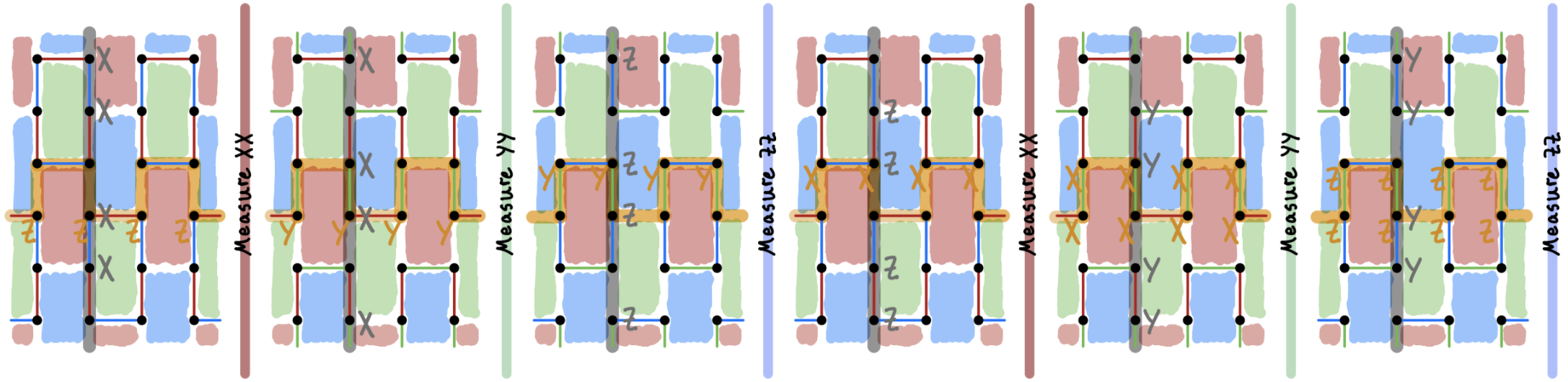}
    }
    \caption[LoF entry]{
        \textbf{The honeycomb code.}
        The layout is a straightened-out hexagonal tiling of the torus with faces 3-colored red (Pauli X), green (Pauli Y), and blue (Pauli Z).  Edges are assigned the color of the faces they span between.
        There is a data qubit at each vertex and, optionally, a measurement ancilla qubit at the center of each edge.
        Each edge represents a 2-body measurement and each face represents a 6-body parity check (stabilizer).  Note that the parity checks commute with all of the edge operators, and so are preserved when measuring them.  As a static subsystem code, this construction protects no degrees of freedom.
        
        The code progresses in repeating rounds, with each round made up of three sub-rounds.  In each sub-round, all the edge parities of one type are measured: first red (X), then green (Y), then blue (Z).  The 6-body parity checks are assembled as the product of the six edges that form their perimeter (measured in consecutive sub-rounds).
        
        The logical qubit's two anti-commuting observables travel along paths highlighted in black and orange.
        As the edges along each observable's path are measured, those measurement results are multiplied into the observable to prevent it from anti-commuting with the next sub-round's edge measurements.
        Consequently, the specific observables along each path change from sub-round to sub-round, cycling with a period of six sub-rounds (two full rounds).
        Note that there is a second pair of nonequivalent anti-commuting logical observables, defined along the same paths but offset by three sub-rounds, making for a total of two encoded logical qubits.  For simplicity, in our analysis, we only consider preserving one of these two.
    }
    \label{fig:honeycomb_code}
\end{figure}

Fortunately, recent work by Hastings \& Haah has alleviated this issue \cite{hastings2021dynamically}.  Rather than defining a static subsystem code from the honeycomb model, their construction manifests `dynamic' logical qubits.  These qubits are encoded into pairs of macroscopic anticommuting observables that necessarily change in time.  Paired with fault-tolerant initialization and measurement, this construction - which Hastings \& Haah  call the \emph{honeycomb code} - serves as an exciting new candidate for a robust error-corrected quantum memory.  A locally equivalent description of this dynamic memory is summarized in \fig{honeycomb_code}. In that layout, a distance-$d$ honeycomb code can be realized as a $1.5d\times d$ periodic lattice of data qubits for $d$ divisible by four.

\subsection{Summary of Results}

While it is helpful to focus on thresholds - the \emph{quality} of physical qubit required to scale a quantum code - we are increasingly interested in the precise \emph{quantity} of physical qubits required to hit target error rates.  In this work, we run numerical simulations to estimate both thresholds and qubit counts required for finite-sized calculations at scale using the honeycomb memory (and for comparison, the surface code).

When entanglement is generated using two-qubit unitary gates, we find honeycomb code thresholds of $0.1\%-0.3\%$, depending on the specifics of the error model. This typically represents approximately half the threshold of the surface code, which we run in the same model (see \fig{threshold}).

We emphasize that, while two-local sparsely-connected codes can be desirable in hardware (e.g. to reduce crosstalk \cite{chamberland2020topological}), two-locality often greatly diminishes the quality of error-correction \cite{bacon2006operator, suchara2011constructions}.  By comparison, our results show that the honeycomb memory remains surprisingly robust, and further, embeds in a lattice of average degree $12/5$.  We study its ``teraquop regime'' - informally, the ratio of physical qubits to logical qubits required to implement trillions of logical operations successfully (see \sec{figures_of_merit}).  Without considering its potential hardware advantages, the honeycomb memory requires between a $5\times$$-$$10\times$ qubit overhead to reach the teraquop regime compared to the surface code at $10^{-3}$ error rates.  However, this gap widens in a superconducting-inspired model in which the measurement channel is the dominant error, as each parity check is assembled as the product of twelve measurements (compared with two in the surface code).

\begin{figure}
    \centering
    \resizebox{0.99\linewidth}{!}{
        \includegraphics{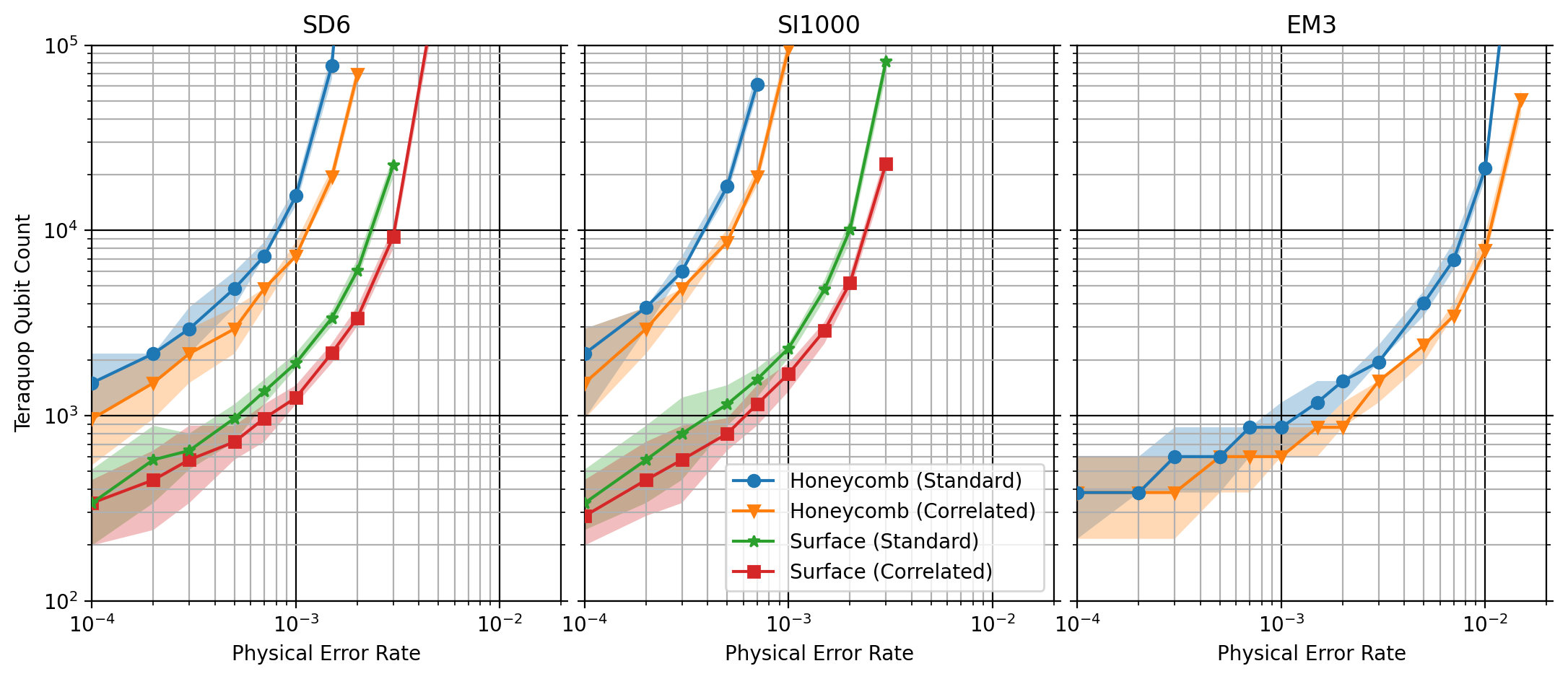}
    }
    \caption{
        \textbf{Teraquop plots} showing the physical qubits per logical qubit required to reach the teraquop regime (one-in-a-trillion per-logical-operation error rates) using both standard (uncorrelated) and correlated minimum-weight perfect-matching.
        SD6 is a standard circuit error model, SI1000 is a superconducting-inspired circuit error model, and EM3 is a circuit error-model with primitive two-body measurements (see \sec{error_models}).
        These values are extrapolated from the line fits in \fig{linefits}.
        Highlighted regions correspond to values that the underlying line fit can be forced to imply while increasing its sum of squares error by at most one (in the natural basis).
        Some discretization effects are visible due to the gaps between achievable code distances.
    }
    \label{fig:teraquop}
\end{figure}

The picture changes qualitatively if entanglement is instead generated using native two-body measurements - a context for which this code is optimized \cite{hastings2021dynamically}.  Direct two-body measurements have been proposed and demonstrated experimentally in circuit QED \cite{chow2010detecting, PhysRevLett.102.200402, DiCarlo_2009, kerckhoff2009physical, Rist__2012, Lalumi_re_2010, DiVincenzo_2013, govia2015scalable, tornberg2014stochastic, criger2016multi, ciani2017three, huembeli2017towards, royer2018qubit, mohseninia2020always, livingston2021experimental}, and are expected to be the native operation in Majorana-based architectures \cite{knapp2018modeling, chao2020optimization, hastings2021dynamically}.  In this setting, we observe a honeycomb code threshold between $1.5\%$ and $2.0\%$ - by comparison, previous work reported a surface code threshold of $0.237\%$ in the same error model \cite{chao2020optimization} (but using a less accurate union-find decoder).  Because of the proximity to the surface code's threshold, the comparative physical qubit savings of a teraquop honeycomb memory at a $10^{-3}$ error rate is several orders of magnitude.

The reason for this reversal of relative performance is that the two-locality of the honeycomb code allows us to jointly measure the data qubits directly.  This significantly reduces the number of noisy operations in a syndrome cycle compared with decomposing the four-body measurements of the surface code, and gracefully eliminates the need for ancilla qubits.  As a result, with two-body measurements at a physical error rate of $10^{-3}$, we project that a teraquop honeycomb memory requires only $600$ qubits (see \fig{teraquop}).  In \sec{conclusion}, we remark on even further reductions to the teraquop qubit count of the honeycomb code that we did not consider.

\section{Error Models}
\label{sec:error_models}

\begin{figure}[htb!]
    \centering
    \resizebox{0.99\linewidth}{!}{
        \includegraphics{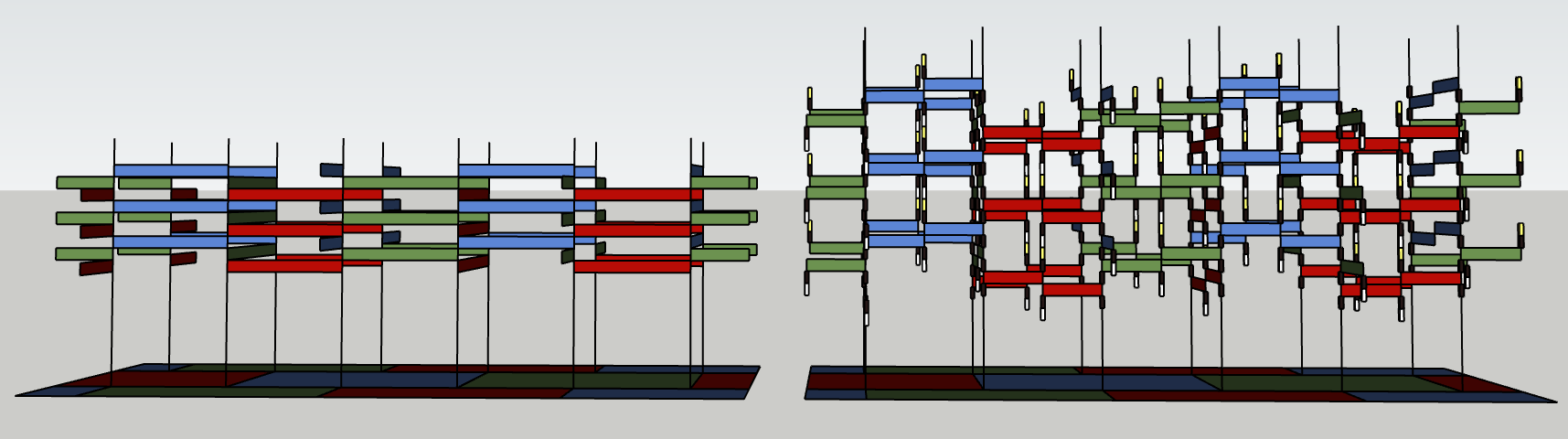}
    }
    \caption{
        Pipelining in honeycomb circuits.
        \textbf{Left}: 3 rounds (9 sub-rounds) of the EM3 honeycomb circuit, which does not require pipelining.
        Vertical flat red/green/blue rectangles are $XX$/$YY$/$ZZ$ parity measurements.
        Vertical black poles represent data qubits.
        \textbf{Right}: 3 rounds (9 sub-rounds) of the SD6 honeycomb circuit, which is pipelined (meaning operations from different rounds and sub-rounds occur during the same time step) to reduce depth.
        Vertical flat red/green/blue rectangles are CNOT gates targeting the measurement ancilla at the center of an edge, with the color indicating the current transformed basis of the data qubit.
        Vertical black poles are data qubits.
        White, yellow, and black boxes are reset operations, measurement operations, and $C_{ZYX}$ operations (the Clifford operation that rotates -120 degrees around $X+Y+Z$) respectively.
        A sketchup file containing this 3d model is attached to the paper as an ancillary file named ``\texttt{3d\_order.skp}''.
    }
    \label{fig:circuits}
\end{figure}

We consider three noisy gate sets, each with an associated circuit-level error model controlled by a single error parameter $p$.  The syndrome extraction sub-cycles are summarized diagrammatically in \fig{circuits}.

\captionsetup[table]{name=Tables}
\begin{table}
    \centering
    \resizebox{\linewidth}{!}{
    \begin{tabular}{|c|l|}
         \hline
         \textbf{Noisy Gate} & \textbf{Definition} \\
         \hline
         $\text{AnyClifford}_2(p)$ & \text{Any 2-qubit Clifford gate, followed by a 2-qubit depolarizing channel of strength $p$.} \\
         \hline
         $\text{AnyClifford}_1(p)$ & Any 1-qubit Clifford gate, followed by a 1-qubit depolarizing channel of strength $p$. \\
         \hline
         $\text{Init}_Z(p)$ & Initialize the qubit as $\ket{0}$, followed by a bitflip channel of strength $p$. \\
         \hline
         $M_Z(p)$ & Precede with a bitflip channel of strength $p$, and measure the qubit in the $Z$-basis. \\
         \hline
          & Measure a Pauli product $PP$ on a pair of qubits and, with probability $p$, choose an error
          \\ $M_{PP}(p)$ & uniformly from the set $\{I,X,Y,Z\}^{\otimes 2} \times \{\text{flip, no flip}\}$.
          \\&The flip error component causes the wrong result to be reported.
          \\&The Pauli error components are applied to the target qubits after the measurement.\\
         \hline
         $\text{Idle}(p)$ & If the qubit is not used in this time step, apply a 1-qubit depolarizing channel of strength $p$. \\
         \hline
         $\text{ResonatorIdle}(p)$ & If the qubit is not measured or reset in a time step during which other qubits are \\ &  being measured or reset, apply a 1-qubit depolarizing channel of strength $p$. \\
         \hline
    \end{tabular}
    }
    \caption{
        Noise channels and the rules used to apply them.  Noisy rules stack with each other - for example, Idle($p$) and ResonatorIdle($p$) can both apply depolarizing channels in the same time step.
        Note that our direct measurement error model is not phenomenological - correlated errors are applied according to the support of each two-qubit measurement, coinciding with the error model in \cite{chao2020optimization}. 
    }
    \label{tbl:noise}
\end{table}
\captionsetup[table]{name=Table}

\begin{table}
    \centering
    \begin{tabular}{|r|l|l|l|}
        \hline
        \textbf{Abbreviation}
        &SD6
        &SI1000
        &EM3
        \\\hline
        \textbf{Name}
            & \begin{tabular}{@{}l@{}}Standard\\Depolarizing\end{tabular}
            & \begin{tabular}{@{}l@{}}Superconducting\\Inspired\end{tabular}
            & \begin{tabular}{@{}l@{}}Entangling\\Measurements\end{tabular}
        \\\hline
        \textbf{Noisy Gateset}
            &\noindent\begin{tabular}{@{}l@{}}
                $\text{CX}(p)$\\
                $\text{AnyClifford}_1(p)$\\
                $\text{Init}_Z(p)$\\
                $M_Z(p)$\\
                $\text{Idle}(p)$\\
            \end{tabular}
            &\begin{tabular}{@{}l@{}}
                \vspace{-0.25cm}
                {} \\
                $\text{CZ}(p)$\\
                $\text{AnyClifford}_1(p/10)$\\
                $\text{Init}_Z(2p)$\\
                $M_Z(5p)$\\
                $\text{Idle}(p/10)$\\
                $\text{ResonatorIdle}(2p)$\\
            \end{tabular}
            &\begin{tabular}{@{}l@{}}
                $\text{M}_{PP}(p)$\\
                $\text{AnyClifford}_1(p)$\\
                $\text{Init}_Z(p)$\\
                $M_Z(p)$\\
                $\text{Idle}(p)$\\
            \end{tabular}
        \\\hline
        \begin{tabular}{@{}r@{}}\textbf{Measurement}\\\textbf{Ancillae}\end{tabular}
            &Yes
            &Yes
            &No
        \\\hline
        \begin{tabular}{@{}r@{}}\textbf{Honeycomb}\\\textbf{Cycle Length}\end{tabular}
            & 6 time steps
            & 7 time steps ($\approx1000$ns)
            & 3 time steps
        \\\hline
    \end{tabular}
    \caption{
        The three noisy gate sets investigated in this paper.
        See \tbl{noise} for the exact definitions of the noisy gates.  The superconducting-inspired acronym refers to an expected cycle time of about $1000$ nanoseconds \cite{chen2021exponential}.
        Note that we still call SD6 ``SD6'' when applying it to the surface code, even though for this model the surface code cycle takes 8 time steps instead of 6.
    }
    \label{tbl:models}
\end{table}

The standard depolarizing model was chosen to compare to the wider quantum error-correction literature.
The entangling measurements model was chosen to represent hardware in which the primitive entangling operation is a measurement, as is expected e.g. in Majoranas \cite{knapp2018modeling, hastings2021dynamically}.
The superconducting-inspired model was chosen to be representative of how we conceive of error contributions on Google's superconducting quantum chip, where measurement and reset take an order of magnitude longer than other gates \cite{chen2021exponential}.

\subsection{Figures of merit}
\label{sec:figures_of_merit}

For each of these error models, we consider three figures of merit - the threshold, the lambda factor, and the teraquop qubit count - each of which depend on the code, decoder, and error model.  Of these three, the threshold is the least indicative of overall performance.  It is the physical error parameter $p_\text{thr}$ below which, for any target logical error rate, there is some (potentially very large) number of qubits we can use to hit that target. 

A better metric is the lambda factor, which additionally depends on a physical error rate $p$.  The lambda factor, $\Lambda(p)$, is the (roughly distance-independent) error suppression obtained by increasing the code distance by two, namely $\Lambda(p) = p_L^d(p)/p_L^{d+2}(p)$.  The lambda factor is related to the threshold as $\Lambda(p_{\text{thr}}) \approx 1$.  The lambda factor in a superconducting quantum repetition code was carefully studied in \cite{chen2021exponential}.

Finally, the most important metric is the so-called teraquop (trillion-quantum-operations) qubit count.  The teraquop qubit count can be extrapolated from the lambda factor, and takes into account the total number of qubits required to realize a particular code.  In particular, at a physical error rate $p$, the teraquop qubit count $T(p)$ is the number of physical qubits required to successfully realize one trillion logical idle gates in expectation (where the idle gate is treated as $d$ rounds, the fundamental building block of a topological spacetime circuit).  Put another way, this metric sets a target physical qubit to logical qubit ratio for realizing a $10^{12}$ gate circuit volume.  For example, with $\sim 10^3 \cdot T(p)$ physical qubits, we could reliably implement a $\sim 10^3$ logical qubit quantum circuit with $\sim 10^9$ layers.  We choose this particular size simply because we believe that it is representative of the computational difficulty of solving important problems such as chemistry simulation \cite{PRXQuantum.2.030305} and cryptography \cite{gidney2021factor,soeken2020improved} using a quantum computer.

\section{Methods}

In order to simulate the honeycomb code, we combine two pieces of software.  The first is Stim, a high performance stabilizer simulator that supports additional error-correction features \cite{gidney2021stim}.  The second is an in-house minimum-weight perfect-matching decoder that uses correlations to enhance performance \cite{fowler2013optimal}.

For each code and error model, we generate a Stim circuit file \cite{stimcircuitformat} describing the stabilizer circuit realizing a memory experiment.
The circuit file contains annotations asserting that certain measurement sets have deterministic parities in the absence of noise (we call these measurement sets \emph{detectors}), as well as annotations stating which measurements are combined to retrieve the final values of logical observables.
Error channels are also annotated into the circuit, with associated probabilities.

Stim is capable of converting a circuit file into a detector error model file \cite{stimdemformat}, which is easier for decoders to consume.
Stim does this by analyzing each error mechanism in the circuit, determining which detectors the error violates and which logical observables the error flips.
We refer to the detectors violated by a single error as the error's \emph{detection event set} or as the error's \emph{symptoms}, and to the logical observables flipped by an error as the error's \emph{frame changes}.
To get a ``graph-like'' error model, Stim uses automated heuristics to split each error's symptoms into a combination of single symptoms and paired symptoms that appear on their own elsewhere in the model.
Stim declares failure if it cannot perform the decomposition step, or if any annotated detector or logical observable is not actually deterministic under noiseless execution.

\begin{figure}[htb!]
    \centering
    \resizebox{0.99\linewidth}{!}{
        \includegraphics{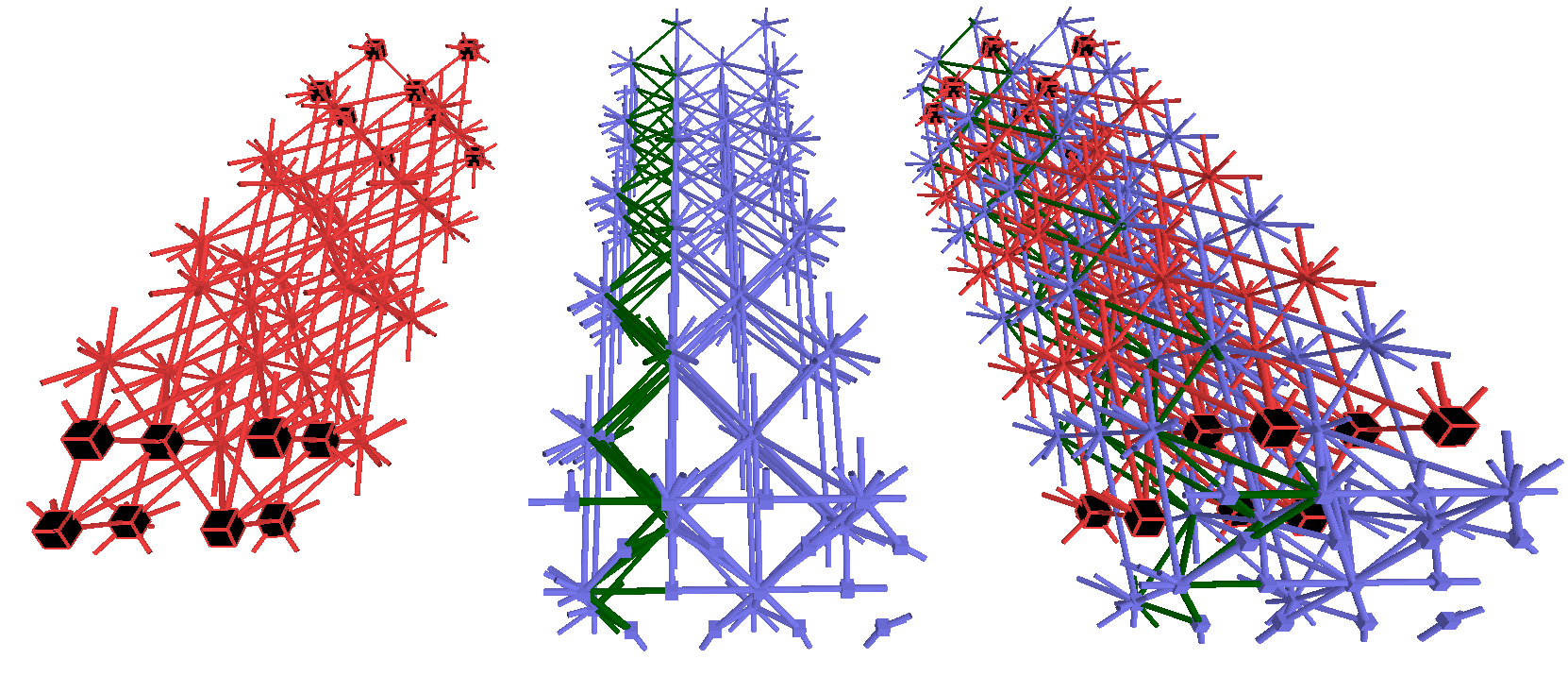}
    }
    \caption{
        The matching graph, computed automatically by Stim, for the 10-round 4x6-data-qubit SD6 honeycomb code circuit initializing and measuring the vertical observable.
        Nodes correspond to detectors (potential detection events) declared in the circuit.
        Edges correspond to error mechanisms that set off two detection events.
        Edges wrapping around due to the periodic boundary conditions are shown terminating in empty space.
        Nodes flipped by an error mechanism that sets off exactly one detection event are indicated by showing the associated node as a large black box.  Due to the periodic spatial boundaries, these only occur at the time boundaries of the matching graph that's missing initialization and terminal measurement detectors.
        Note that for an error to wrap around the lattice yielding a logical failure, at least four edges must be traversed, consistent with the expected effective distance of 4.
        \textbf{Left}: The connected component of the matching graph that corrects errors that commute with the observable we're preserving.
        \textbf{Center}: The connected component of the matching graph that corrects errors that anti-commute with the observable we're preserving.
        Edges highlighted in green correspond to errors that flip the specific logical observable annotated in the circuit.
        \textbf{Right}: The complete matching graph.
        The graph has degree at most 12, similar to the surface code.
        By contrast, the EM3 honeycomb circuits produce matching graphs with nodes of degree 18 due to additional correlated errors.
    }
    \label{fig:matchgraph}
\end{figure}

The decoder takes the graph-like detector error model and converts it into a weighted matching graph with edges and boundary edges corresponding to the decomposed two- and one-element detection event sets.
See \fig{matchgraph} for an example of a honeycomb code matching graph.
The decoder optionally notes how errors with more than two symptoms have been decomposed into multiple graph-like (edge-like) errors, and derives reweighting rules between the edges corresponding to the graph-like pieces.
These reweighting rules are used for correlated decoding.

Stim is also capable of sampling the circuit at a high rate, producing detection events which are passed to the decoder.
Having initialized into the $+1$-eigenspace of an observable, we declare success when the decoder correctly guesses the logical observable's measurement outcome based on the detection event data provided to it.

\section{Results and Discussion}

\begin{figure}[htb!]
    \centering
    \resizebox{\linewidth}{!}{
        \includegraphics{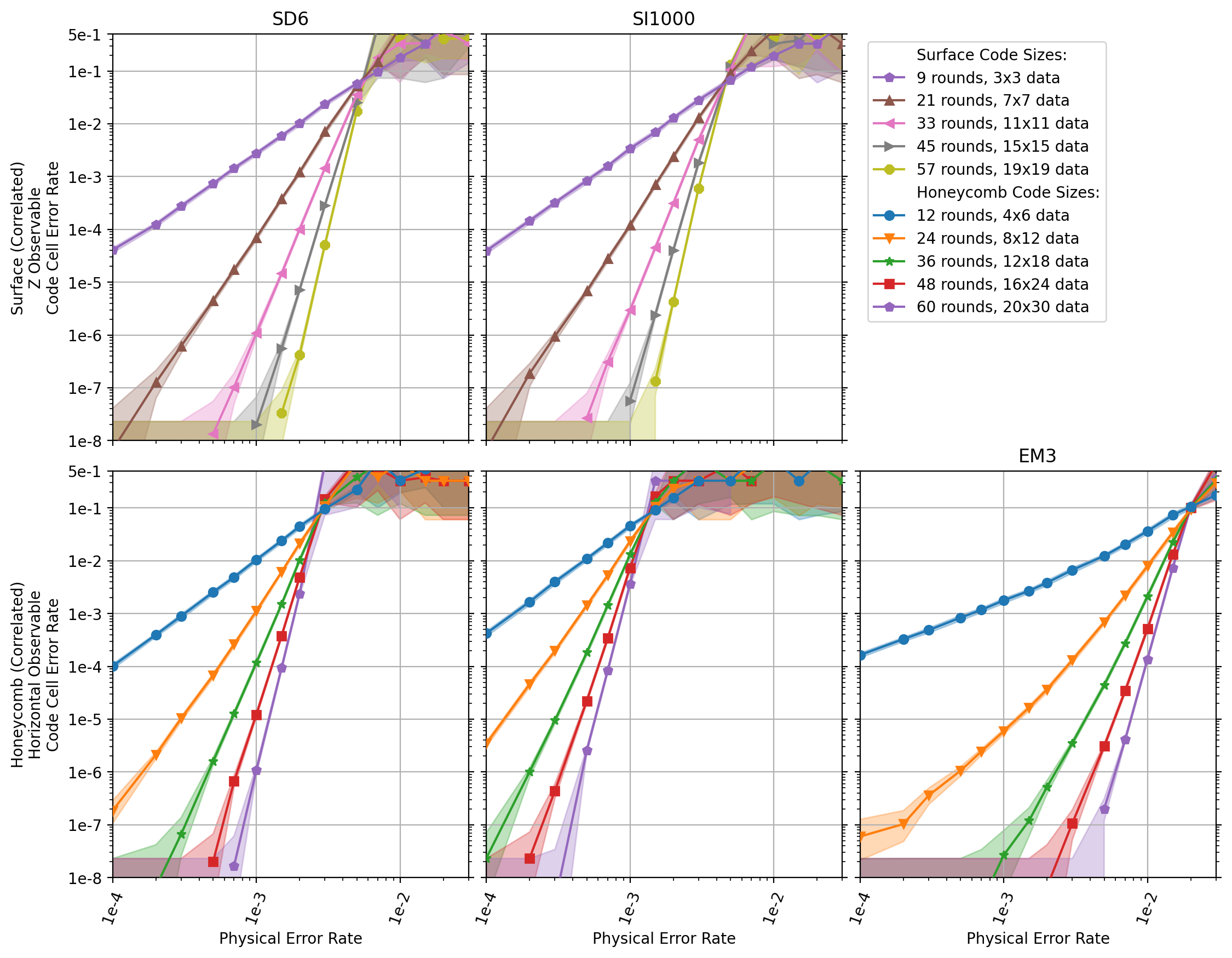}
    }
    \caption{
        \textbf{Threshold plots} showing failure rates per code cell for one of the observables in the honeycomb code and the surface code under various error models.
        For more cases, see \fig{threshold_all} in the appendix.
        By ``code cell'', we mean a spacetime region with a spacelike extent that realizes a code distance of $d$, and a timelike extent of $d$ rounds.
        We simulate $3d$ rounds (but report the error rate per $d$ rounds) to minimize time-boundary effects. 
        Our honeycomb code description on a torus only realizes distances that are multiples of four.
        Highlighted areas correspond to hypotheses whose likelihoods are within a factor of 1000 of the maximum likelihood estimate.
        The large highlighted areas above threshold are due to aggressive termination of sampling when logical error rates near 50\% are detected, resulting in less samples.
    }
    \label{fig:threshold}
\end{figure}

\begin{figure}[htb!]
    \centering
    \resizebox{\linewidth}{!}{
        \includegraphics{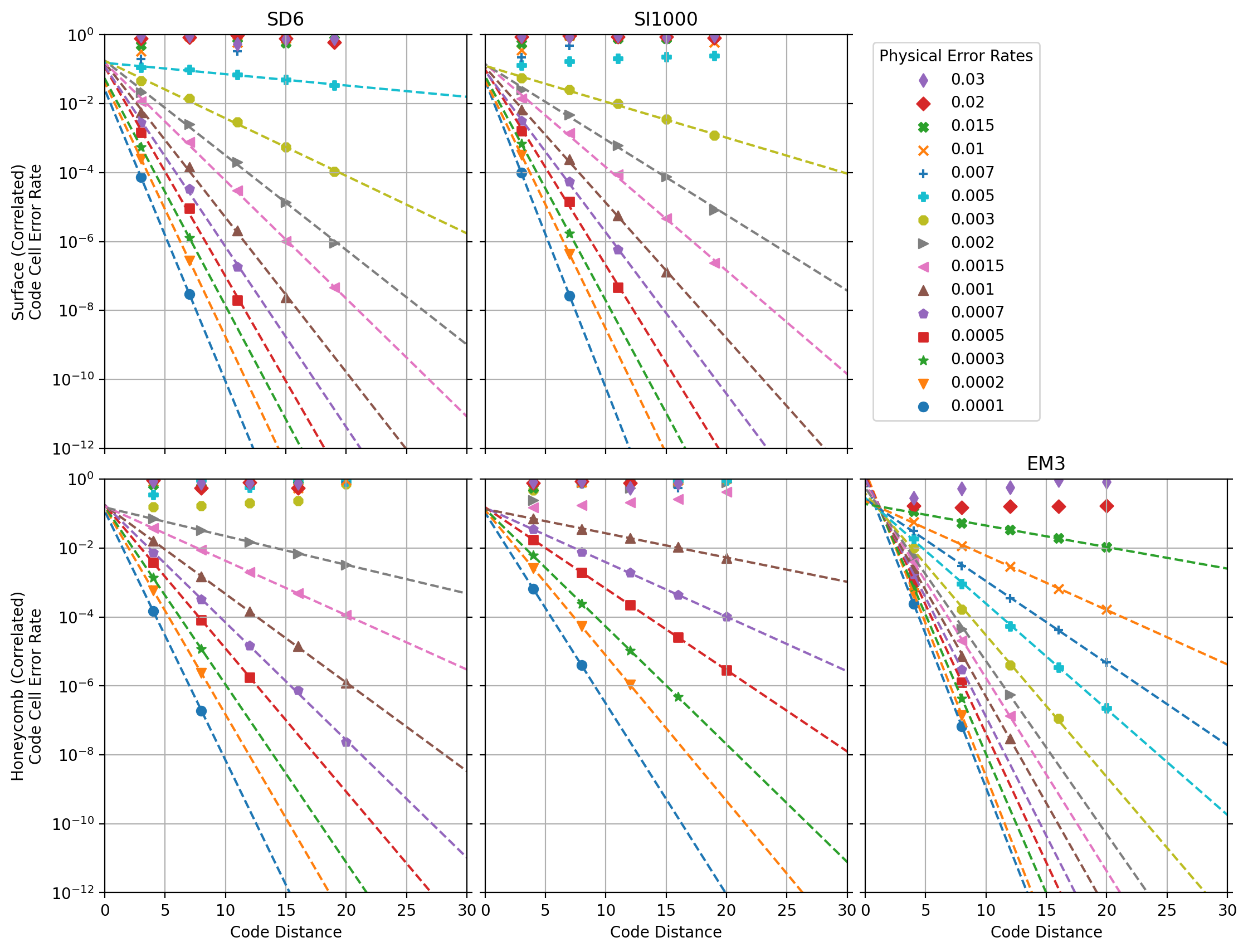}
    }
    \caption{
        \textbf{Line fit plots} projecting logical error rate as a function of code distance at various error rates.
        For more cases, see \fig{linefits_all} in the appendix.
        This plot combines the logical error rates of both observables by assuming an error occurs if either fails independently, leading to a slight upper bound on the error.
        When a noise rate is above threshold (i.e. the logical error rate increases with code distance), no line fit is shown.
        The line fits in this figure are used to generate the lambda plot (\fig{lambda}) and the teraquop plot (\fig{teraquop}).
    }
    \label{fig:linefits}
\end{figure}

Logical error rates are presented in \fig{threshold} and \fig{linefits}.  
Note that we report the logical error rate per distance-$d$-round block, which can depress the threshold and lambda estimate relative to reporting the logical error rate per cycle \cite{stephens2014fault}.
In the standard circuit model SD6, we observe a threshold within $0.2\%-0.3\%$ for the honeycomb code, which compares with a surface code threshold within $0.5\%-0.7\%$.  The relative performance of the honeycomb code is reduced in the SI1000 model, wherein we observe a threshold of $0.1\%-0.15\%$ compared to a surface code threshold of $0.3\%-0.5\%$.  This lower relative performance may be attributed to the proportionally noisier measurement channel.  Within the honeycomb code, detectors are assembled as products of twelve individual measurements.  When measurements are the dominant error channel in the circuit, these twelve-measurement detectors carry significant noise compared to the two-measurement detectors in the surface code.

The EM3 error model flips the story.
There, we observe a threshold of within $1.5\%-2\%$ for the honeycomb code.
Unlike SD6 and SI1000, we elected not to simulate the surface code in this model, as there are several avenues for optimization already explored in previous work \cite{chao2020optimization}.
In an identical error model, they report a threshold of $0.237\%$, significantly below the honeycomb code.
Unfortunately, the two numbers are \emph{not} directly comparable - in \cite{chao2020optimization}, a union-find decoder was used, which can result in lower thresholds.
However, as the union-find decoder is moderately accurate by comparison \cite{huang2020fault}, we do not expect this difference to change qualitatively.

The very high EM3 honeycomb code threshold stems from three factors.  First, two-local codes benefit greatly from a two-body measurement circuit architecture. Because of their two-locality, we can measure the data qubits directly without decomposing the effective measurement into a product of noisy gates.  The result is a circuit with significantly less overall noise.  Second, the particular error model we adopted from \cite{chao2020optimization} is somewhat atypical.  The collective failure rate of the two-body measurement includes both the noise it introduces to data qubits as well as the accuracy of its measurement - a demanding metric.  While this does induce a richer correlated error model, it introduces less overall noise than two error channels applied independently to the qubit support and measurement outcome (see \fig{detectionfraction} in the appendix).  However, we can still infer the excellent performance of the honeycomb code by comparing it to the surface code - we make such comparisons to avoid differences in absolute performance that depend on error model details \cite{stephens2014fault}. Third, the honeycomb code is simply an excellent two-local code.  While many other two-local codes involve high-weight parity checks \cite{bacon2006operator, suchara2011constructions}, the honeycomb supports parity checks of weight six.  As a result, these checks are less noisy and the resulting code is more robust.

It is worth noting that both codes appear to achieve the full effective code distance in both the SD6 and SI1000 models.  However, the honeycomb code does not achieve the full distance in the EM3 model.  This is because many of the two-body measurements fully overlap with the observable, and so the effective distance of the code is halved due to correlated errors arising from these direct measurements.

However, the inclusion of ancilla qubits in the EM3 model would create two deleterious effects.  First, it multiplies the total number of qubits at a particular code distance by $2.5\times$.  Consequently, we only require $1.6\times$ more qubits to achieve a particular effective distance without ancilla, rather than the $4\times$ overhead required when only counting data qubits.  Second, the inclusion of ancilla qubits generates significant overhead in syndrome circuit complexity.  While increasing the effective distance is guaranteed to be more efficient in the low-$p$ limit, it appears that entropic effects contribute more in the $10^{-4}-10^{-2}$ physical error regime.  This can also be inferred from the curve of the logical error rates, as $p \rightarrow 10^{-4}$ shows the relative contribution of these low-weight failure mechanisms beginning to dominate the more numerous higher-weight failure mechanisms.  Overall, it appears the trade-off is well worth it.

\begin{figure}[htb!]
    \centering
    \resizebox{0.99\linewidth}{!}{
        \includegraphics{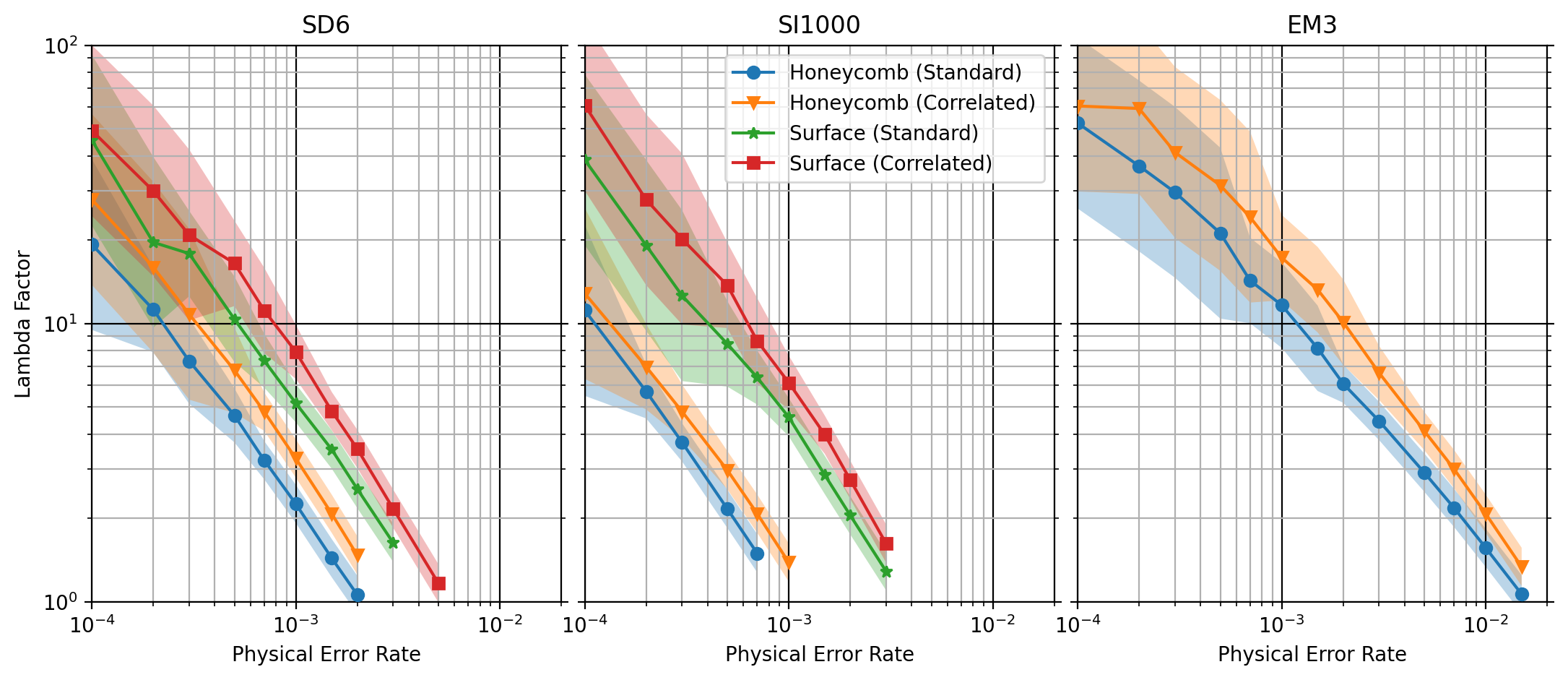}
    }
    \caption{
        \textbf{Lambda plots} showing the average factor by which the per-logical-operation error rate is suppressed when increasing the code distance by 2 (extrapolated from line fits in \fig{linefits}).
        Because we increase the honeycomb's code distance in jumps of 4, these values are the square roots of the minimum increase in that case.
        Highlighted regions correspond to values that the underlying line fit can be forced to imply while increasing its sum of squares error by at most one (in the natural basis).
    }
    \label{fig:lambda}
\end{figure}

\fig{lambda} presents the error suppression factor lambda using both correlated and standard (uncorrelated) decoding.  Here, it is clear that the surface code offers greater error suppression in SD6 and SI1000, and that including correlations in the decoder boosts this error suppression.  This latter point is important because including correlations does \emph{not} significantly augment the threshold - however, it is this below-threshold error suppression that we are most interested in.

Finally, \fig{teraquop} presents the teraquop regime for each code.  This combines the lambda factor for each code with the total number of qubits required to realize that error suppression.  At distance-$d$ under an independent depolarizing model, the rotated surface code requires $2d^2 - 1$ data qubits in total.  This compares with either $1.5d^2$ qubits without ancillae, or $3.75d^2$ qubit with ancillae in the honeycomb code - although it is worth noting that, while we protect only one logical qubit, two are encoded.

In line with the threshold and lambda estimates, we observe that the surface code is significantly more qubit efficient in SD6 and SI1000 at all error rates.  However, we observe that in EM3, at a physical error rate of $10^{-3}$, we can achieve a teraquop memory using only $600$ qubits.

It is important to note the shape of these curves.  In particular, there is a large savings in teraquop qubit count as the error rate diminishes \emph{near threshold}.  However, far below threshold, these gains taper off.  Phrased in terms of lambda, it is extremely beneficial to operate at an error rate with a lambda factor of $10$ rather than $2$, but achieving a lambda factor of $50$ rather than $10$ is far less important.
    
\section{Conclusions}
\label{sec:conclusion}

In this work, we numerically tested the robustness of the honeycomb code and compared it with the surface code.  We found that the honeycomb code is exceptionally robust among two-local codes in entangling unitary circuit models, and exceptionally robust among all codes when assuming primitive two-body measurements.  In presenting these results, we introduced the notion of the teraquop qubit count, which we believe is a good quantification of a code's overall finite-size performance.  For comparison, we tested the surface code in two of the three models to provide baseline estimates for its teraquop regime as well.

Very recently, a planar version of the honeycomb code \emph{with} boundary was proposed \cite{haah2021boundaries}.  Technical differences including modifications to the gate sequence, new error paths terminating at boundaries, and performance during logical gates more make this planar variant worthy of additional study. 
Note that the honeycomb code's estimated teraquop qubit count is sensitive to changes in the boundary conditions.
For example, the periodic boundaries currently allow for two logical qubits to be stored instead of one.\footnote{The same holds true for the rotated toric code \cite{bombin2007optimal}, although the halved distance of one of the two logical qubits (due to hook errors) makes this benefit less clear-cut.}
However, in this work we did not count the second logical qubit, as we are unsure how viable using both logical qubits would be in the context of a complete fault-tolerant computational system.

Also, shortly before completing this work, we realized that rotating/shearing the periodic layout~\cite{bombin2007optimal} would use $25\%$ fewer physical qubits to achieve the same code distance.
However, we have not simulated this layout, and this change may affect the lambda factor of the code.
Of course, multiplying the estimated teraquop qubit count (at a $10^{-3}$ physical error rate) of $600$ by $1/2$ (for the second logical qubit) and $3/4$ (for the rotated/sheared layout) yields an encouraging number, but because of the preceding caveats, we do not do so here.

\section{Author Contributions}

Craig Gidney wrote code defining, verifying, simulating, and plotting results from the various honeycomb code circuits.
Michael Newman reviewed the implementation of the code and the simulation results, generated the surface code circuits to compare against, and handled most of the paper writing.
Austin Fowler wrote the high performance minimum-weight perfect-matching decoder used by the simulations and helped check simulation results.
Michael Broughton helped scale the simulations over many machines.

\section{Acknowledgements}

We thank Jimmy Chen for consulting on reasonable parameters to use for the superconducting-inspired error model.
We thank Rui Chao for helping us understand the details of the EM3 error model, and for helpful discussions on the gap in performance between the surface code and the honeycomb code in this model.
We thank Cody Jones for conversations on multi-qubit measurement and for feedback on an earlier draft.
We thank Dave Bacon and Ken Brown for feedback on an earlier draft.
We thank Twitter user @croller for suggesting the name ``teraquop regime''.
Finally, we thank Hartmut Neven for creating an environment in which this work was possible.

\bibliographystyle{plainnat}
\bibliography{refs}
\clearpage
\appendix

\section{Detection Event Fractions}

\begin{figure}[htb!]
    \centering
    \resizebox{0.99\linewidth}{!}{
        \includegraphics{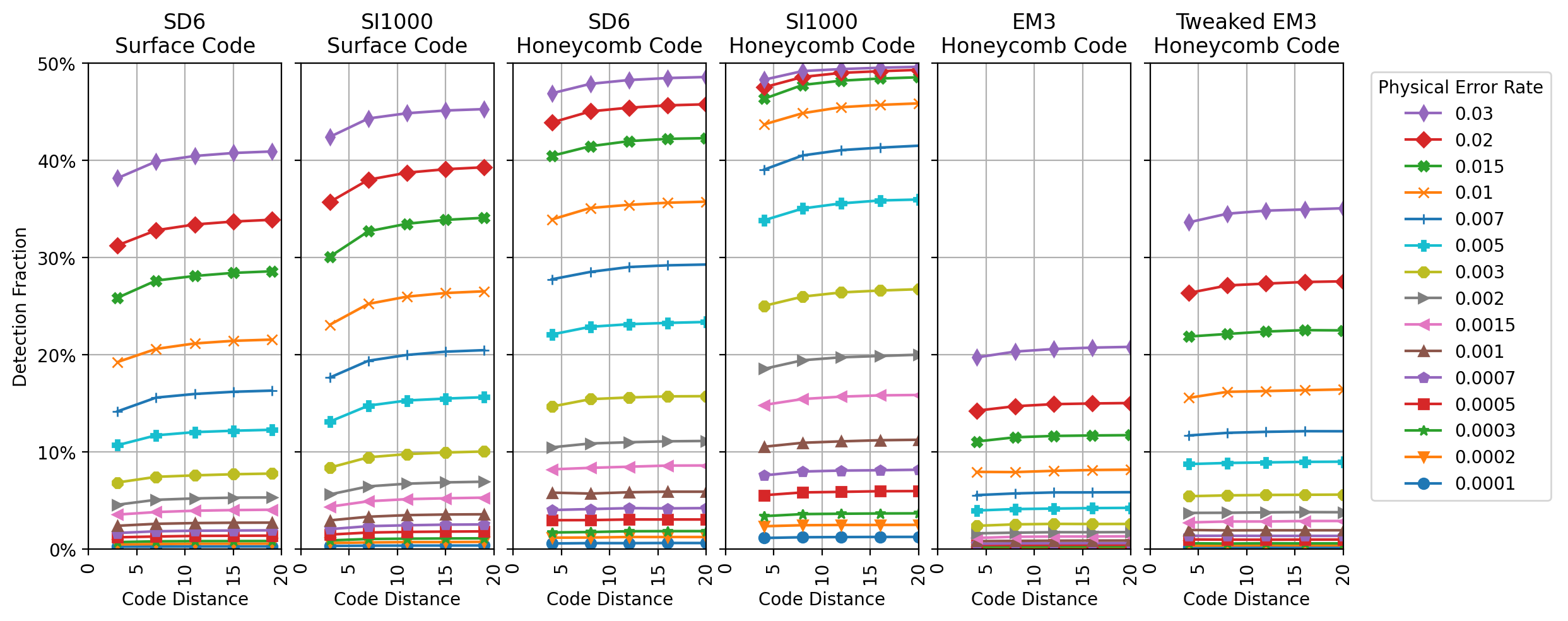}
    }
    \caption{
        \textbf{Detection fraction plots} showing the average probability of a detector producing a detection event during simulated runs for various circuits, sizes, and noise parameters.  Note that there is a slight initial uptick in detection event fraction, which quickly stabilizes at higher distances.  This is due to nearby initialization and terminal measurement phases, which tend to be less noisy.
    }
    \label{fig:detectionfraction}
\end{figure}

One measure of the amount of noise in a circuit is the detection event fraction, namely, the proportion of actual detection events to the total number of potential detection events.  In \fig{detectionfraction}, we plot the detection event fraction for each combination of error model and code.  Notably, the EM3 error model has a detection fraction significantly lower than the other models in the honeycomb code.
We compared a tweaked EM3 variant that separates the error due to the projective action of the measurement on the data qubits from the error on the measurement outcome.
The tweaked EM3 model replaces the 32-case correlated error that occurs with probability $p$ during a parity measurement with two independent errors: a measurement-result-flipping error which occurs with probability $p$ and a two qubit depolarizing error with strength $p$ applied to the qubits before the measurement.
This tweaked model gives a detection fraction that appears closer to other error models, and was in fact used in \cite{bravyi2012subsystem}.  This suggests that grouping the measurement and data errors into a combined error budget may be overoptimistic.  In particular, if these errors occurred independently, they would each have to have probability $\approx p/2$ to match the entropy in the EM3 model (although the EM3 model produces more correlated errors).  With that said, we expect that even in the EM3 model, the detection event fraction for the surface code remains quite high \cite{chao2020optimization}.
Of course, the various error models are coarse approximations of quantum noise under different (and sometimes projected) architectures. Consequently, we do not have the context to set the scale of the parameters used in the EM3 model, and so we caution against comparing different models directly.

\section{Converting Disjoint Error Mechanisms into Independent Error Mechanisms}

The parity measurement error mechanism used for the EM3 error model is defined in terms of disjoint errors.
Error models produced by Stim can only use independent error mechanisms.
This mismatch is handled by converting the disjoint cases into independent cases, using the formula

$$p_{\text{ind}} = \frac{1}{2} - \frac{1}{2} \sqrt[{2^{n-1}}]{1 - p_{\text{mix}}}$$

where $n$ is the number of basis errors (e.g. we use $n=5$ because our basis errors are $X_1$, $Z_1$, $X_2$, $Z_2$, $\text{flip}$), $p_{\text{mix}}$ is the probability of choosing to apply one of the component errors in a maximally mixing channel, and  $p_{\text{ind}}$ is the probability of applying each error independently so that the concatenation of the independent channels produces the same error distribution as the maximally-mixing channel \cite{chao2020optimization}.

\section{Additional Data}

To avoid overloading the reader, figures in the main text of the paper show representative cases for the data that we collected.
For example, some figures show data from runs protecting one observable but not from runs protecting the other, and some figures show error rates when using correlated matching but not when using standard matching.
This appendix section includes more exhaustive figures, showing data on all of the cases we tried.

We have also made the data we collected available in CSV format, attached to this paper as an ancillary file named \path{honeycomb_memory_stats.csv}.
Each row in the CSV file summarizes how many shots were taken and how many errors were seen for a particular code distance, error model, decoding strategy, etc.

For each case we considered, we took shots until reaching 100 million shots or 1000 errors, or seeing clear evidence of being above threshold (logical error rate near 50\%); whichever came first.
Often we took slightly more shots than needed because runs were batched and distributed over multiple machines.
In total, we spent roughly 10 CPU core years on the simulations.

All of the data and figures in the paper were generated using python code available online at \url{https://github.com/strilanc/honeycomb_threshold}.
The python code is also attached to the paper as ancillary files with paths beginning with \path{repo/}.

Correlated matching was performed with an internal tool that is not part of the repository.
This makes it more difficult to reproduce the data in this paper that is based on correlated matching, because correlated matching involves hand-tuned heuristics (although implementing the decoder in \cite{fowler2013optimal} would be a good start).
It is more feasible to reproduce the data based on standard matching since the minimum-weight perfect-matching problem has well-defined solutions that other decoders should agree on up to degeneracy (e.g. our code includes functionality for using PyMatching \cite{higgott2021pymatching} to decode).

\begin{figure}[htb!]
    \centering
    \resizebox{0.99\linewidth}{!}{
        \includegraphics{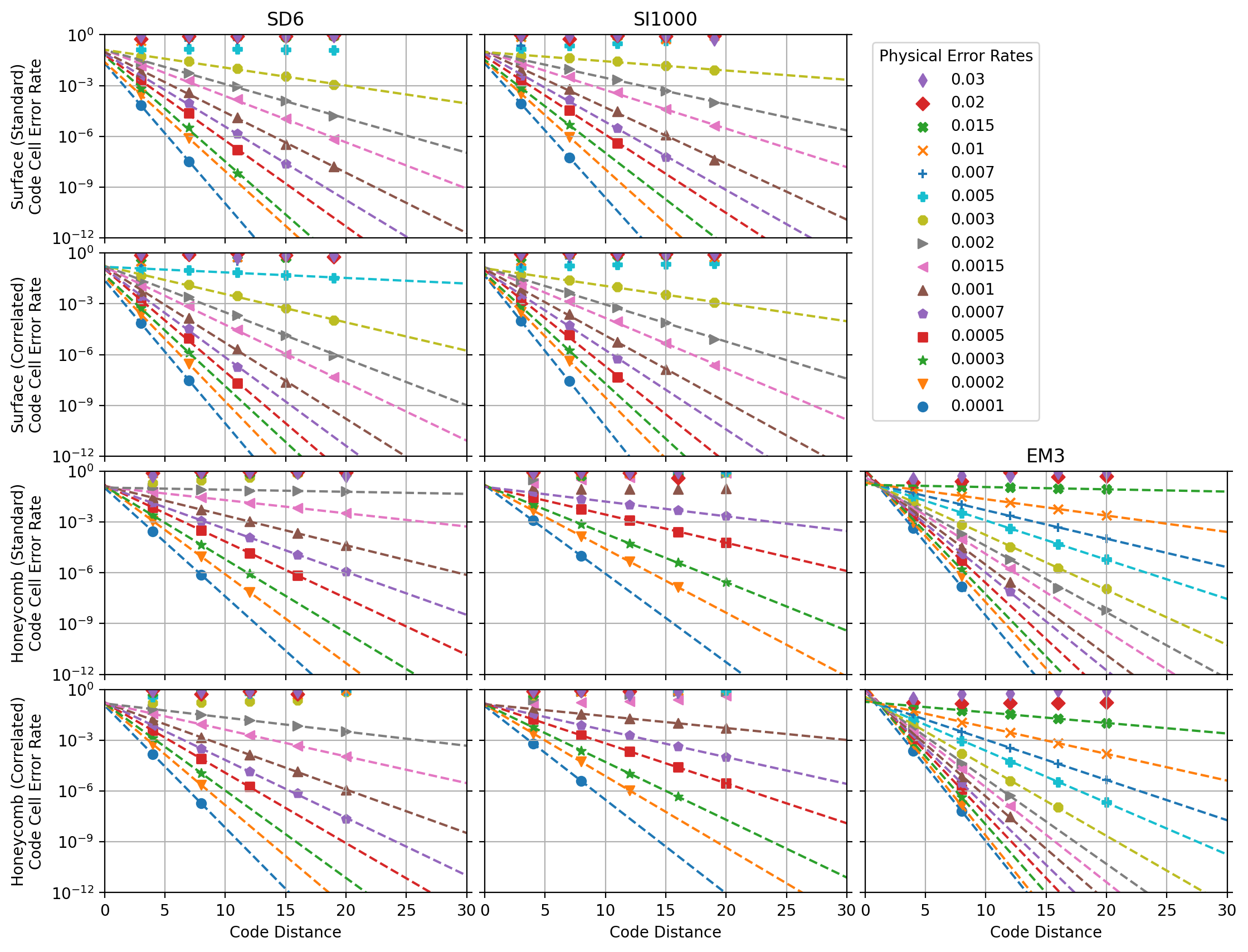}
    }
    \caption{
        \textbf{Exhaustive line fit plots} projecting logical error rates as code distance is increased at different error rates for the honeycomb code and the surface code using various noisy gate sets and either standard or correlated decoding.
    }
    \label{fig:linefits_all}
\end{figure}

\begin{figure}[htb!]
    \centering
    \resizebox{!}{500pt}{
        \includegraphics{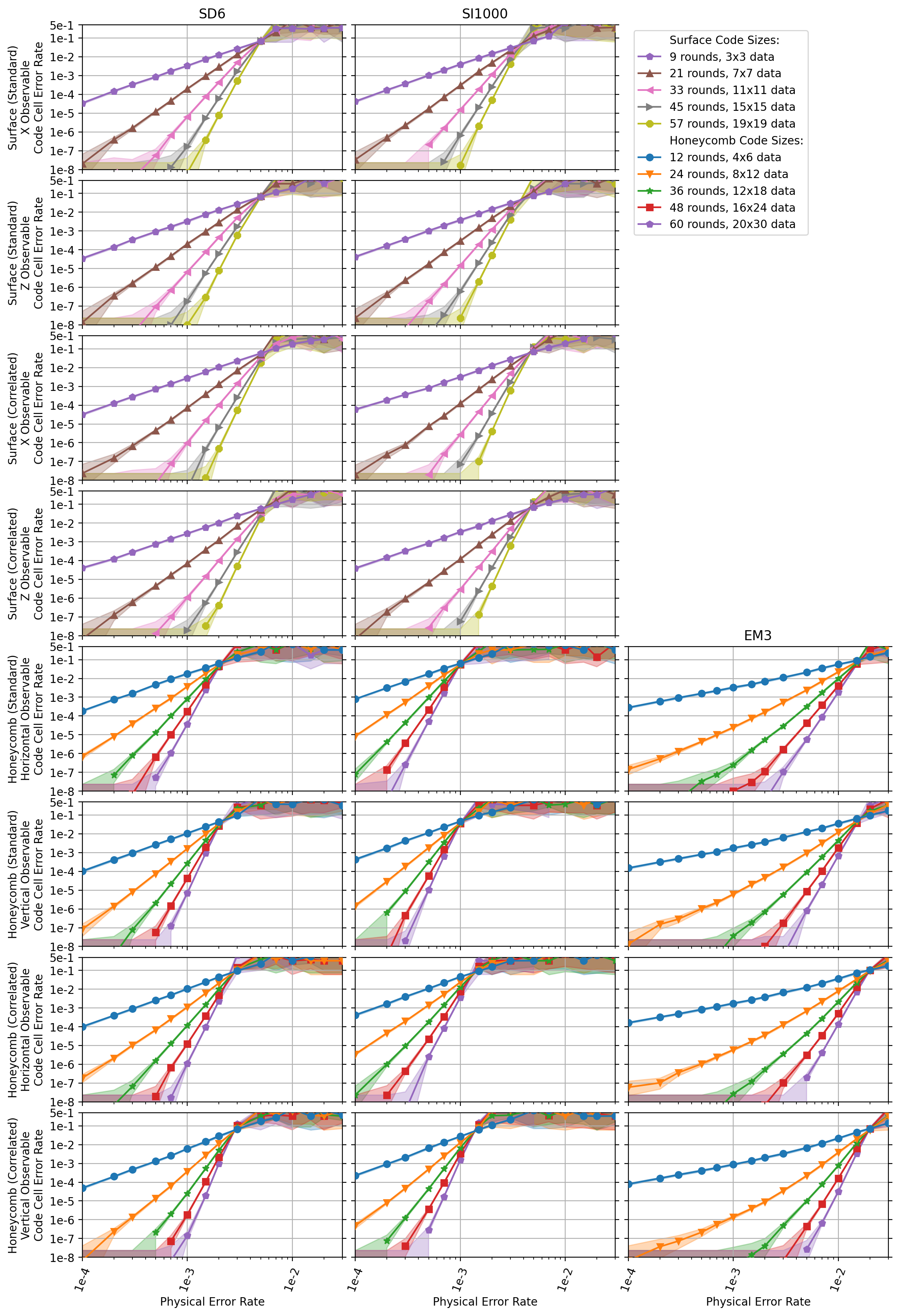}
    }
    \caption{
        \textbf{Exhaustive threshold plots} showing error rates per code cell for both observables of the honeycomb code and the surface code using various noisy gate sets and either standard or correlated decoding.
    }
    \label{fig:threshold_all}
\end{figure}

\section{Example Honeycomb Circuit}

The following is a Stim circuit file \cite{stimcircuitformat} describing a honeycomb memory experiment with 2$\times$6 data qubits and 335 rounds.
The last round is terminated early, before its second sub-round.
The circuit fault-tolerantly initializes the horizontal observable, preserves it through noise, and then fault-tolerantly measures it.

The circuit includes coordinate annotations, noise annotations, and detector/observable annotations.
The noise model is the ``tweaked EM3" noise model mentioned in \fig{detectionfraction} because that noise model produces the simplest circuit file.

Because copying code from PDF files is not reliable, this circuit is also attached to the paper as the ancillary file ``\path{example_honeycomb_circuit.stim}".

\begin{lstlisting}[style=stimcircuit]
# Annotate locations of qubits.
QUBIT_COORDS(1, 0) 0
QUBIT_COORDS(1, 1) 1
QUBIT_COORDS(1, 2) 2
QUBIT_COORDS(1, 3) 3
QUBIT_COORDS(1, 4) 4
QUBIT_COORDS(1, 5) 5
QUBIT_COORDS(3, 0) 6
QUBIT_COORDS(3, 1) 7
QUBIT_COORDS(3, 2) 8
QUBIT_COORDS(3, 3) 9
QUBIT_COORDS(3, 4) 10
QUBIT_COORDS(3, 5) 11

# Initialize data qubits into |+> state.
R 0 1 2 3 4 5 6 7 8 9 10 11
X_ERROR(0.001) 0 1 2 3 4 5 6 7 8 9 10 11
TICK
H 0 1 2 3 4 5 6 7 8 9 10 11
DEPOLARIZE1(0.001) 0 1 2 3 4 5 6 7 8 9 10 11
TICK

# X sub-round. Compare X parities to X initializations.
DEPOLARIZE2(0.001) 9 3 2 1 4 5 0 6 7 8 11 10
MPP(0.001) X9*X3 X2*X1 X4*X5 X0*X6 X7*X8 X11*X10
OBSERVABLE_INCLUDE(1) rec[-3]
DETECTOR(0, 3, 0) rec[-6]
DETECTOR(1, 1.5, 0) rec[-5]
DETECTOR(1, 4.5, 0) rec[-4]
DETECTOR(2, 0, 0) rec[-3]
DETECTOR(3, 1.5, 0) rec[-2]
DETECTOR(3, 4.5, 0) rec[-1]
SHIFT_COORDS(0, 0, 1)
TICK

# Y sub-round. Get X*Y=Z stabilizers for first time.
DEPOLARIZE2(0.001) 7 1 2 3 0 5 4 10 9 8 11 6
MPP(0.001) Y7*Y1 Y2*Y3 Y0*Y5 Y4*Y10 Y9*Y8 Y11*Y6
OBSERVABLE_INCLUDE(1) rec[-6]
SHIFT_COORDS(0, 0, 1)
TICK

# Z sub-round. Get Y*Z=X stabilizers to compare against initialization.
DEPOLARIZE2(0.001) 11 5 0 1 4 3 2 8 7 6 9 10
MPP(0.001) Z11*Z5 Z0*Z1 Z4*Z3 Z2*Z8 Z7*Z6 Z9*Z10
OBSERVABLE_INCLUDE(1) rec[-5] rec[-2]
DETECTOR(0, 0, 0) rec[-12] rec[-10] rec[-7] rec[-6] rec[-5] rec[-2]
DETECTOR(2, 3, 0) rec[-11] rec[-9] rec[-8] rec[-4] rec[-3] rec[-1]
SHIFT_COORDS(0, 0, 1)
TICK

# X sub-round. Get Z*X=Y stabilizers for the first time.
DEPOLARIZE2(0.001) 9 3 2 1 4 5 0 6 7 8 11 10
MPP(0.001) X9*X3 X2*X1 X4*X5 X0*X6 X7*X8 X11*X10
OBSERVABLE_INCLUDE(1) rec[-3]
SHIFT_COORDS(0, 0, 1)
TICK

# Stable state reached. Can now consistently compare to stabilizers from previous round.
REPEAT 333 {
    # Y sub-round. Get X*Y = Z stabilizers to compare against last time.
    DEPOLARIZE2(0.001) 7 1 2 3 0 5 4 10 9 8 11 6
    MPP(0.001) Y7*Y1 Y2*Y3 Y0*Y5 Y4*Y10 Y9*Y8 Y11*Y6
    OBSERVABLE_INCLUDE(1) rec[-6]
    DETECTOR(0, 2, 0) rec[-30] rec[-29] rec[-26] rec[-24] rec[-23] rec[-20] rec[-12] rec[-11] rec[-8] rec[-6] rec[-5] rec[-2]
    DETECTOR(2, 5, 0) rec[-28] rec[-27] rec[-25] rec[-22] rec[-21] rec[-19] rec[-10] rec[-9] rec[-7] rec[-4] rec[-3] rec[-1]
    SHIFT_COORDS(0, 0, 1)
    TICK

    # Z sub-round. Get Y*Z = X stabilizers to compare against last time.
    DEPOLARIZE2(0.001) 11 5 0 1 4 3 2 8 7 6 9 10
    MPP(0.001) Z11*Z5 Z0*Z1 Z4*Z3 Z2*Z8 Z7*Z6 Z9*Z10
    OBSERVABLE_INCLUDE(1) rec[-5] rec[-2]
    DETECTOR(0, 0, 0) rec[-30] rec[-28] rec[-25] rec[-24] rec[-23] rec[-20] rec[-12] rec[-10] rec[-7] rec[-6] rec[-5] rec[-2]
    DETECTOR(2, 3, 0) rec[-29] rec[-27] rec[-26] rec[-22] rec[-21] rec[-19] rec[-11] rec[-9] rec[-8] rec[-4] rec[-3] rec[-1]
    SHIFT_COORDS(0, 0, 1)
    TICK

    # X sub-round. Get Z*X = Y stabilizers to compare against last time.
    DEPOLARIZE2(0.001) 9 3 2 1 4 5 0 6 7 8 11 10
    MPP(0.001) X9*X3 X2*X1 X4*X5 X0*X6 X7*X8 X11*X10
    OBSERVABLE_INCLUDE(1) rec[-3]
    DETECTOR(0, 4, 0) rec[-30] rec[-28] rec[-25] rec[-24] rec[-22] rec[-19] rec[-12] rec[-10] rec[-7] rec[-6] rec[-4] rec[-1]
    DETECTOR(2, 1, 0) rec[-29] rec[-27] rec[-26] rec[-23] rec[-21] rec[-20] rec[-11] rec[-9] rec[-8] rec[-5] rec[-3] rec[-2]
    SHIFT_COORDS(0, 0, 1)
    TICK
}

# Transversal measurement.
H 0 1 2 3 4 5 6 7 8 9 10 11
DEPOLARIZE1(0.001) 0 1 2 3 4 5 6 7 8 9 10 11
TICK
X_ERROR(0.001) 0 1 2 3 4 5 6 7 8 9 10 11
M 0 1 2 3 4 5 6 7 8 9 10 11
OBSERVABLE_INCLUDE(1) rec[-11] rec[-5]

# Compare X data measurements to X parity measurements from last sub-round.
DETECTOR(0, 3, 0) rec[-18] rec[-9] rec[-3]
DETECTOR(1, 1.5, 0) rec[-17] rec[-11] rec[-10]
DETECTOR(1, 4.5, 0) rec[-16] rec[-8] rec[-7]
DETECTOR(2, 0, 0) rec[-15] rec[-12] rec[-6]
DETECTOR(3, 1.5, 0) rec[-14] rec[-5] rec[-4]
DETECTOR(3, 4.5, 0) rec[-13] rec[-2] rec[-1]

# Compare X data measurements to previous X stabilizer reconstruction.
DETECTOR(0, 0, 0) rec[-30] rec[-28] rec[-25] rec[-24] rec[-23] rec[-20] rec[-12] rec[-11] rec[-7] rec[-6] rec[-5] rec[-1]
DETECTOR(2, 3, 0) rec[-29] rec[-27] rec[-26] rec[-22] rec[-21] rec[-19] rec[-10] rec[-9] rec[-8] rec[-4] rec[-3] rec[-2]
\end{lstlisting}

\end{document}